\documentclass[graybox]{svmult}
\usepackage{mathptmx}       % selects Times Roman as basic font
\usepackage{helvet}         % selects Helvetica as sans-serif font
\usepackage{courier}        % selects Courier as typewriter font
\usepackage{type1cm}        % activate if the above 3 fonts are
\usepackage{makeidx}         % allows index generation
\usepackage{graphicx}        % standard LaTeX graphics tool
\usepackage{multicol}        % used for the two-column index
\usepackage[bottom]{footmisc}% places footnotes at page bottom

\usepackage{amssymb,amsmath}

\makeindex             % used for the subject index
\begin{document}

\newcommand{\sn}{{\rm sn}\,}
\newcommand{\cn}{{\rm cn}\,}
\newcommand{\dn}{{\rm dn}\,}

\title*{Analytical solutions for geodesics in black hole spacetimes}
\author{Eva Hackmann and Claus L\"ammerzahl}

\institute{Eva Hackmann \at ZARM, University Bremen, Am Fallturm, 28359 Bremen, Germany, \email{eva.hackmann@zarm.uni-bremen.de}
\and Claus L\"ammerzahl \at ZARM, University Bremen, Am Fallturm, 28359 Bremen, Germany \email{claus.laemmerzahl@zarm.uni-bremen.de}}

\maketitle

\abstract{We review the analytical solution methods for the geodesic equations in Kerr-Newman-Taub-NUT-de Sitter spacetimes and its subclasses in terms of elliptic and hyperelliptic functions. A short guide to corresponding literature for general timelike and lightlike motion is also presented.}

\section{Introduction}
\label{sec:1}

Black holes belong to the most fascinating objects in astrophysics and are well suited to explore the regime of strong gravity. We consider here black holes which are described by the six parameter family of Kerr-Newman-Taub-NUT-de Sitter spacetimes including mass, rotation, electric and magnetic charge, gravitomagnetic mass (or NUT charge), and the cosmological constant. Maybe the best way to explore the gravitational field of such objects is through the observation of the motion of small massive particles and light, which is described by the geodesic equation. The complete set of orbits can best be analyzed using analytical methods. Already in 1931, Hagihara \cite{Hagihara} used Weierstrass elliptic functions to analytically solve the geodesic equation in Schwarzschild spacetime. Later, Darwin \cite{Darwin1959,Darwin1961} solved the geodesic equations in Schwarzschild spacetime in terms of Jacobian elliptic functions. These methods and their generalization to hyperelliptic functions can be used to solve the geodesic equation in the six parameter spacetime under consideration. Although this requires only mathematics from the 19th century, surprisingly the geodesic equations in Schwarzschild-de Sitter spacetime were analytically solved only in 2008 \cite{HackmannLaemmerzahl2008,HackmannLaemmerzahl08b}. Here we will review these solution methods and provide a short literature guide.

\section{The Kerr-Newman-Taub-NUT-de Sitter space-time}\label{sec:spactime}
We consider here spherically or axially symmetric spacetimes with up to six parameters, which are part of the family of type D Pleba\'nski-Demia\'nski spacetimes. The metric is given by \cite{GriffithsPodolsky06,MankoRuiz2005}
\begin{multline} \label{PD_metric}
ds^2 = \frac{\Delta_r}{\rho^2} \left(dt - (a \sin^2\theta  + 2 n \cos\theta) d\varphi\right)^2 - \frac{\rho^2}{\Delta_{r}} dr^2 \\
- \frac{\Delta_{\theta}}{\rho^2} \sin^2\theta (a dt - (r^2 + a^2 + n^2) d\varphi)^2 - \frac{\rho^2}{\Delta_{\theta}} d\theta^2
\end{multline}
with
\begin{align}
\rho^2 & = r^2 + \left(n - a \cos\theta \right)^2 \,, \nonumber \\
\Delta_{\theta} & = 1 + \tfrac{1}{3} a^2 \Lambda \cos^2\theta - \tfrac{4}{3} \Lambda a n \cos\theta\,, \\
\Delta_{r} & = %\left( 1- \frac{\Lambda}{3} r^2 - \Lambda n^2 \right) (r^2+a^2-n^2) - 2Mr + Q_{\rm e}^2 + Q_{\rm m}^2 - \frac{4}{3} \Lambda n^2 r^2 \nonumber
r^2  - 2 M r + a^2 - n^2 + Q^2_{\rm e} + Q_{\rm m}^2 - \tfrac{1}{3} \Lambda \left(r^4 + (6 n^2 + a^2) r^2 + 3 (a^2 - n^2) n^2\right)\,,\nonumber
\end{align}
where $M$ is the mass, $a=J/M$ the specific angular momentum, $\Lambda$ the cosmological constant, $n$ is the gravitomagnetic mass, $Q_{\rm e}$ is the electric, and $Q_{\rm m}$ the magnetic charge of a gravitating source.

\section{Analytical solution methods}\label{sec:AS}
The motion of test particles in the spacetime metric \eqref{PD_metric} is given by the geodesic equation
\begin{align}
\frac{d^2x^\mu}{ds^2} + \Gamma^\mu{}_{\nu\rho} \frac{dx^\nu}{ds} \frac{dx^\rho}{ds} = 0
\end{align}
where $(\Gamma^\mu{}_{\nu\rho})$ are the Christoffel symbols and $s$ is an affine parameter. As the metric \eqref{PD_metric} is axially symmetric there exists the two constants of motion
\begin{align}
E & = u_\mu \xi^\mu_{(t)}\,, \quad L_z = - u_\mu \xi^\mu_{(\varphi)}\,. \label{genEL}
\end{align}
connected to the Killing vectors $\xi_{(t)}$ and $\xi_{(\varphi)}$, which can be interpreted as the energy and the specific angular momentum in direction of the symmetry axes. Here $u$ denotes the four-velocity. If we have even spherical symmetry, i.e.~for $a=n=0$, these two constants of motion together with the restriction to the equatorial plane, which is without loss of generality, and the normalization $g_{\mu\nu}dx^\mu dx^\nu = \epsilon$, where $\epsilon=0$ for light and $\epsilon=1$ for massive particles, is sufficient to separate the geodesic equation. This yields then the differential equation
\begin{align}
\left(\frac{dr}{ds}\right)^2 & = E^2-\frac{\Delta_r}{r^2} \left(\epsilon + \frac{L^2}{r^2}\right)\,,
\end{align}
and the energy and angular momentum take the form
\begin{align}
E=g_{tt}\frac{dt}{ds}\,, \quad L=L_z=r^2\frac{d\varphi}{ds}\,.
\end{align}
If the spacetime is axially symmetric, we cannot in general restrict the motion of test particles to the equatorial plane. Therefore, we need an additional constant of motion. In 1968 Carter \cite{Carter68} surprisingly found such a constant of motion, which can be derived as a separation constant and is not connected to an obvious symmetry of the spacetime. With this constant the geodesic equation can again be separated and we get equations of motions in the form  
\begin{align}
\rho^4 \left( \frac{dr}{ds}\right)^2 & = ((r^2+a^2+n^2)E-aL_z)^2 - \Delta_r(\epsilon r^2+K)\,, \label{axial:drds}\\
\rho^4 \left(\frac{d\theta}{ds}\right)^2 & = \Delta_\theta (K-\epsilon(n-a\cos\theta)^2) - \frac{(E(a\sin^2\theta+2n\cos\theta)-L_z)^2}{\sin^2\theta}\,, \label{axial:dthetads}
\end{align}
where $K$ is the Carter constant. If $n=0$ then $K=(aE-L_z)^2$ corresponds to motion restricted to the equatorial plane. From the expression of energy and angular momentum \eqref{genEL} we get the additional equations
\begin{align}
\rho^2 \frac{d\varphi}{ds} & = \frac{a}{\Delta_r} ((r^2+a^2+n^2)E-aL_z) - \frac{E(2n\cos\theta+a\sin^2\theta)-L_z}{\Delta_\theta \sin^2\theta}\,,\\
\rho^2 \frac{dt}{ds} & = \frac{r^2+a^2+n^2}{\Delta_r} ((r^2+a^2+n^2)E-aL_z)\nonumber\\
& \quad - \frac{a\sin^2\theta+2n\cos\theta}{\Delta_\theta\sin^2\theta} (E(2n\cos\theta+a\sin^2\theta)-L_z) \,.
\end{align}
Note that the equations \eqref{axial:drds} and \eqref{axial:dthetads} are still coupled by the factor $\rho^2$. This issue was solved by Mino in 2003 \cite{Mino03} by introducing a new affine parameter $\lambda$ defined by $\frac{ds}{d\lambda}=\rho^2$.%,
% \begin{align}
% \left( \frac{dr}{d\lambda}\right)^2 & = \frac{a}{\Delta_r} ((r^2+a^2+n^2)E-aL_z) - \frac{E(2n\cos\theta+a\sin^2\theta)-L_z}{\Delta_\theta \sin^2\theta} \label{axial:drdl}\\
% \left(\frac{d\theta}{d\lambda}\right)^2 & = \frac{r^2+a^2+n^2}{\Delta_r} ((r^2+a^2+n^2)E-aL_z)\nonumber\\
% & \quad - \frac{a\sin^2\theta+2n\cos\theta}{\Delta_\theta\sin^2\theta} (E(2n\cos\theta+a\sin^2\theta)-L_z) \,. \label{axial:dthetadl}
% \end{align}

If we now consider in the spherically symmetric case the differential equations for $r(\varphi)$,
\begin{align}
\left(\frac{dr}{d\varphi}\right)^2 & = \frac{r^4}{L^2} \left( E^2-\frac{\Delta_r}{r^2} \left(\epsilon + \frac{L^2}{r^2}\right) \right)
\end{align}
we see that on the right hand side we have a polynomial of degree three or four, if the cosmological constant vanishes, and of degree five or six in general. The same holds for the differential equations \eqref{axial:drds} and \eqref{axial:dthetads}. This kind of differential equations can be solved in terms of elliptic functions if the polynomial is of degree three or four, and in terms of hyperelliptic functions if it is of degree five or six.

\subsection{Solutions in terms of elliptic functions}
For differential equations of the general type
\begin{align}
\left(\frac{dx}{dy}\right)^2 & = P_{3,4}(x)\,, \quad x(y_0)=x_0\label{genellipticdiff}
\end{align}
where $P_{3,4}$ is a polynomial of degree three or four, there are basically two (equivalent) solution methods based on the Jacobian elliptic function $\sn$ and on the Weierstrass elliptic function $\wp$. The first can be defined as the inverse of an elliptic integral,
\begin{align}
z = \int_0^w \frac{dt}{\sqrt{(1-t^2)(1-k^2t^2)}} \quad \Rightarrow \quad \sn(z;k)=w
\end{align}
where $0\leq k\leq 1$ is the modulus, $w\in [0,1]$, and $z \in \mathbb{R}$. The Weierstrass elliptic function is given as a series
\begin{align}
\wp(z;2\omega_1,2\omega_2) & = \frac{1}{z^2} + \sum_{\omega_{nm} \in \Omega} \left( \frac{1}{(z-\omega_{nm})^2} - \frac{1}{\omega_{nm}^2} \right)\,,
\end{align}
where $2\omega_1$,$2\omega_2$ are the periods of $\wp$ ($\frac{\omega_1}{\omega_2} \notin \mathbb{R}$) and $\Omega=\{ \omega_{nm} \in \mathbb{C}| \omega_{nm}=2n\omega_1+2m\omega_2\}$. It solves the initial value problem (see e.g.~\cite{Markushevich77, Hurwitz})
\begin{align}
\left(\frac{dx}{dy}\right)^2 = 4x^3-g_2x-g_3\,, \quad x(0)=\infty \label{Weierstrassform}
\end{align}
where $g_2 = 60 \sum_{\omega_{nm}\in \Omega} \omega_{nm}^{-4}$, $g_3 = 140 \sum_{\omega_{nm}\in \Omega} \omega_{nm}^{-6}$. Note that both $\sn$ and $\wp$ can be written in terms of the Riemann theta function
\begin{align}
\theta[\tau v+w](z;\tau) & = \sum_{m \in \mathbb{Z}^g} \exp( \pi i (m+v)^t (\tau (m+v) + 2z + 2w) )\,, \label{deftau}
\end{align}
where $z\in\mathbb{C}^g$, $\tau$ is a $g\times g$ symmetric matrix with positive definite imaginary part, $\tau v+w \in \mathbb{C}^g$ is the characteristic, and $g$ is the genus, here $g=1$. %Then 
% \begin{align}
% \sn(z;k) = k^{-\frac{1}{2}} \frac{\theta[\frac{1}{2}](\frac{z}{2K(k)};\tau)}{\theta[\frac{\tau+1}{2}](\frac{z}{2K(k)};\tau)}\,, \quad \wp(z;\omega_1,\omega_2) = \frac{1}{(2\omega_1)^2} \wp\left(\frac{z}{2\omega_1};1,\frac{\omega_2}{\omega_1}\right) = 
% \end{align}
% where $K(k)$ is the complete elliptic integral of the first kind and $\tau=\frac{iK(\sqrt{1-k^2})}{K(k)}$, and $\wp=...$.

The general differential equation \eqref{genellipticdiff} can be solved in terms of Jacobian elliptic functions by applying a substitution which converts the problem to the form
\begin{align}
\left(\frac{d\tilde{x}}{dy}\right)^2 & = (1-\tilde{x}^2)(1-k^2\tilde{x}^2)\,.
\end{align}
This substitution depends on the degree of $P_{3,4}$ and its number of complex zeros as well as on the type of orbit you want to obtain. For a list see e.g.~\cite{Abramowitz}.

To find a solution of \eqref{genellipticdiff} in terms of the Weierstrass elliptic function you have to convert the problem to the standard form \eqref{Weierstrassform}, which can be obtained by first converting a polynomial of degree four to degree three by $x=\xi^{-1}+x_P$ if $P_{3,4}=\sum_i a_ix^i$ and $P_{3,4}(x_P)=0$ and subsequently, or if $P_{3,4}$ was of degree three in the first place, substituting $\xi=\frac{1}{b_3}(4z+\frac{b_2}{3})$ if $P=\sum_i b_ix^i$ is the polynomial of degree three.

Note that you always have to take care of the initial condition, too.

\subsection{Solutions in terms of hyperelliptic functions}
To generalize the solution methods outlined in the previous section we first need to consider the Jacobi inversion problem 
\begin{align}
y_i = \sum_{j=1}^g \int_{\infty}^{x_j} \frac{t^{i-1}dt}{\sqrt{P(t)}}\,, \quad i=1,\ldots,g \label{Jacobiinversion}
\end{align}
where $P(t)=4t^{2g+1}+\sum_{n=0}^{2g} a_nt^n$ is a polynomial of degree $2g+1$ and $g$ is the genus. Note that for $g=1$ and $a_2=0$ we recover \eqref{Weierstrassform}. The $g$ solutions $x_j(y_1,\ldots,y_g)$ of \eqref{Jacobiinversion} can be given in terms of generalized Weierstrass functions. These are defined by the theta function via the Kleinian sigma function
\begin{align}
\sigma(z;\omega_1,\omega_2) & = C e^{iz^t\kappa z} \theta[K_\infty](z;\omega_1^{-1}\omega_2)\,,\\
\wp_{ij}(z;\omega_1,\omega_2) & = - \frac{\partial}{\partial z_i} \frac{\partial}{\partial z_j} \log \sigma(z;\omega_1,\omega_2)\,,
\end{align}
where $z\in \mathbb{C}^g$, $\omega_i$ are $g \times g$ matrices such that $\omega_1^{-1}\omega_2$ is symmetric with positive definite imaginary part, $\kappa = \eta(2\omega_1)^{-1}$ with the periods of the second kind $2\eta$, $K_\infty$ is the vector of Riemann constants, and $C$ is a constant which does not matter here (for further details see e.g.~\cite{BuchstaberEnolskiiLeykin97}; note that $K_\infty=\tau(\frac{1}{2},\frac{1}{2})^t + (0,\frac{1}{2})^t$ if $g=2$). The solutions of \eqref{Jacobiinversion} are then given by the solutions of
\begin{align}
x^g + \sum_{i=1}^g \wp_{gi}(y_1,\ldots,y_g) x^{i-1} = 0\,.
\end{align}
Let us consider now the case $g=2$ and a general differential equation of the form 
\begin{align}
\left(\frac{dx}{dy}\right)^2 & = P_{5,6}(x)\,, \quad x(y_0)=x_0\label{genhyperdiff}
\end{align}
where $P_{5,6}$ is a polynomial of degree five or six. If it is of degree six it can be reformulated as $\tilde{x} \frac{d\tilde{x}}{dy}=\sqrt{P_5(\tilde{x})}$ by a substitution $x=\tilde{x}^{-1}+x_P$, where $x_P$ is a zero of $P_{5,6}$ and $P_5$ is a polynomial of degree five. Such a differential equation, or, if $P_{5,6}$ was of degree five in the first place, can then be cast in the form 
\begin{align}
t^{i-1}\frac{dt}{dy}=\sqrt{P(t)}\,, \quad i=1,2 \label{normhyperdiff}
\end{align}
by an appropriate normalization. We may then find the solution to this equation as the limiting case $x_2\to\infty$ of the Jacobi inversion problem \eqref{Jacobiinversion} in the following way: first observe that 
\begin{align}
t & := x_1 = \lim_{x_2\to\infty} \frac{x_1x_2}{x_1+x_2} = \lim_{x_2\to\infty} \frac{\wp_{12}(y_1,y_2)}{\wp_{22}(y_1,y_2)}\nonumber\\
& = \lim_{x_2\to\infty} \frac{\sigma\sigma_{12}-\sigma_1\sigma_2}{\sigma_2^2-\sigma\sigma_{22}} (y_1,y_2)\,,\label{solx}
\end{align}
where $\sigma_i(z)$ is the derivative of $\sigma$ with respect to $z_i$. From \eqref{Jacobiinversion} with $g=2$ and $x_2\to\infty$ we may identify either $y_1$ or $y_2$ with our physical coordinate $y$ in the differential equation \eqref{normhyperdiff}, say $y_i$. Fortunately, we get automatically rid of the other $y_j$, $j\neq i$, by the same limiting process $x_2\to\infty$. This is because the set of zeros of the theta function $z\mapsto \theta[K_\infty]((2\omega)^{-1} z)$, which is a one dimensional submanifold of $\mathbb{C}^2$, is given by all vectors $z=(z_1,z_2)$ which can be written as $z_i=\int_\infty^x \frac{t^{i-1}dt}{\sqrt{P(t)}}$ with the same $x$ (see e.g.~\cite{Mumford83}). This is exactly true for the vector $(y_1,y_2)$, which means that we may write $y_j=f(y_i)$ for some function $f$. As the zeros of the theta function are also zeros of $\sigma$ we can simplify \eqref{solx} to
\begin{align}
t & = - \frac{\sigma_1}{\sigma_2} (y_1,f(y_1)) \quad \text{ or } \quad t = - \frac{\sigma_1}{\sigma_2} (f(y_2),y_2)\,.
\end{align}
Note that according to the given initial condition we actually have that $y_i$ is the physical coordinate minus a constant.

\section{Analytical solutions in the literature}
In this section we will collect applications of the methods outlined in section \ref{sec:AS} to geodesic motion in the Kerr-Newman-Taub-NUT-de Sitter spacetime as given in section \ref{sec:spactime}. For older literature we refer to Sharp \cite{Sharp1979} who collected most of the papers on geodesic motion in Kerr-Newman spacetime and subclasses, which were available at that time. Partly this is still quite complete but we also try to update his collection (with respect to analytical solutions). Note that we only consider analytical solutions to general timelike and lightlike geodesics (with an electric or magnetic charge, as applicable). In particular, we do not list the vast literature on equatorial motion in Kerr spacetime. Of course, we do not claim that our list is complete.

{\bf{Schwarzschild:}} Regarding analytical solution methods the list of Sharp already contained the complete set of solutions. Most notably, this includes the works by Hagihara \cite{Hagihara}, who derived the analytical solutions in terms of Weierstrass elliptic functions, and Darwin \cite{Darwin1959,Darwin1961}, who used Jacobian elliptic functions and integrals.

{\bf{Reissner-Nordstr\"om:}} Surprisingly, the analytical solutions to the geodesic equation in Reissner-Nordstr\"om spacetime seem to be considered first only in 1983 by Gackstatter \cite{Gackstatter1983} although it can be handled completely analogously to the Schwarzschild case. He studied bound timelike geodesics and light in terms of Jacobian elliptic integrals and functions. Recently, Slez\'{a}kov\'{a} \cite{Slezakova2006} gave a comprehensive analysis of arbitrary timelike, lightlike, and even spacelike geodesics. Grunau and Kagramanova \cite{Grunau2011} solved the equations of motion of electrically and magnetically charged particles in Reissner-Nordstr\"om spacetime in terms of Weierstrass elliptic functions.

{\bf{Taub-NUT:}} Timelike geodesics were studied by Kagramanova et al \cite{Kagramanova2010} in terms of Weierstrass elliptic functions.

{\bf{Kerr:}} Most of the older literature on Kerr spacetime is concerned with the much simpler particular case of equatorial geodesics. We refer to Sharp \cite{Sharp1979} here for these early works. Note that in terms of the proper time (or the corresponding affine parameter for light) the equations of motion are still coupled. Therefore, most of the analytical solutions before the introduction of the Mino time \cite{Mino03} implicitly included integrals over the latitude or the radius, see e.g. Kraniotis \cite{Kraniotis2005} or Slez\'{a}kov\'{a} \cite{Slezakova2006} for a review. As notable exception, \v{C}ade\v{z} et al \cite{Cadezetal1998} introduced already in 1998 a similar parameter (called $P$, see their equation (34)) as they considered the motion of light. 

After the introduction of the Mino time, in 2009 Fujita and Hikida \cite{FujitaHikida09} used this new affine parameter to derive the analytical solution for bound timelike geodesics in terms of Jacobian elliptic functions. General timelike geodesics and lightlike motion were treated shortly after that by Hackmann \cite{HackmannDiss} in 2010.  

Note that Kraniotis \cite{Kraniotis2011} also derived analytical solutions for lightlike geodesics in terms of hypergeometric functions.

{\bf{Kerr-Newman:}} Charged particle motion was considered by Hackmann and Xu \cite{Hackmannetal2013} in terms of Weierstrass elliptic functions.

{\bf{Schwarzschild-de Sitter}}, also called Kottler space-time: Note that on the level of the differential equation, lightlike geodesics in Schwarzschild-de Sitter are identical with the lightlike equations of motion for Schwarzschild, as the cosmological constant can be absorbed in the definition of just a single parameter. Analytical solution are given e.g. in Gibbons et al \cite{Gibbonsetal2008}. General timelike geodesics in Kottler spacetime can be treated in terms of hyperelliptic functions as elaborated by Hackmann and L\"ammerzahl \cite{HackmannLaemmerzahl2008,HackmannLaemmerzahl08b}.

{\bf{Reissner-Nordstr\"om-de Sitter:}} The equations of motion for general timelike geodesics were solved in \cite{Hackmannetal2008}. The motion of photons was very recently analytically calculated by Villanueva et al \cite{Villanuevaetal2013} for a negative cosmological constants using Weierstrass elliptic functions.

{\bf{Taub-NUT-de Sitter:}} Timelike motion was analyzed in \cite{Hackmannetal2009}.

{\bf{Kerr-de Sitter:}} The equations of motions for timelike geodesics were analytically solved by Hackmann et al \cite{Hackmannetal2009,Hackmannetal2010} in terms of hyperelliptic functions. Note that Kraniotis \cite{Kraniotis2011} also derived analytical solutions for lightlike geodesics in terms of hypergeometric functions.

{\bf{Kerr-Newman-Taub-NUT-de Sitter:}} The general solution for timelike geodesic was shortly outlined in \cite{Hackmannetal2009}.

\begin{acknowledgement}
We thank the German Research Foundation DFG for financial support within the Research Training Group 1620 ``Models of Gravity''. 
\end{acknowledgement}

% \bibliographystyle{plain}
% \bibliography{literature_short}

\end{document}